\documentclass[rnote]{aa} 

\usepackage{graphicx}
\usepackage{txfonts}
\begin{document}

   \title{Effects of moderate abundance changes on the atmospheric
          structure and colours of Mira variables}

   \titlerunning{Effects of moderate abundance changes in Miras}
   \authorrunning{M. Scholz et al.}

   \author{M. Scholz
          \inst{1,2}
          \and
           M.J. Ireland 
          \inst{3,4}
          \and
           P.R. Wood 
          \inst{5}
          }

   \institute{
             Zentrum f\"ur Astronomie der Universit\"at Heidelberg (ZAH),
             Institut f\"ur Theoretische Astrophysik, Albert-Ueberle-Str.2,
             69120 Heidelberg, Germany
             \and
             Sydney Institute for Astronomy (SIfA), School of Physics,
             University of Sydney NSW 2006, Australia\\
             \email{michael.scholz@uni-heidelberg.de}
             \and
             Department of Physics and Astronomy, Macquarie University,
             North Ryde NSW 2109, Australia
             \and
             Australian Astronomical Observatory, Epping NSW 1710, Australia\\
             \email{michael.ireland@anu.edu.au}
             \and
             Research School of Astronomy and Astrophysics, Australian National
             University, Canberra ACT 2600, Australia\\
             \email{peter.wood@anu.edu.au}
             }

   \date{Received 04 December 2013 / Accepted 24 April 2014}

 
  \abstract
    {}
    {We study the effects of moderate deviations from solar abundances
    upon the atmospheric structure and colours of typical Mira
    variables.}
    {We present two model series of dynamical opacity-sampling
    models of Mira variables which have
    (1) $\frac{1}{3}$ solar metallicity and
    (2) ``mild'' S-type C/O abundance ratio ([C/O]=0.9) with typical
    Zr enhancement (solar +1.0).
    These series are compared to a previously studied solar-abundance
    series which has similar fundamental parameters (mass, luminosity,
    period, radius) that are close to those of $o$~Cet.}
    {Both series show noticeable effects of abundance upon stratifications
    and infrared colours but cycle-to-cycle differences mask
    these effects at most pulsation phases, with the exception of
    a narrow-water-filter colour near minimum phase.}
    {}

   \keywords{stars: AGB and post-AGB -- stars: atmospheres -- 
             stars: abundances}

   \maketitle
%

\section{Introduction}

  The density stratification of upper atmospheric layers of Mira
variables is determined by outwards travelling shock fronts. These
shock fronts are seen in typical emission lines of hydrogen and
other atoms (Fox et al. 1984; Richter \& Wood 2004)
and lead to a geometrically very extended stellar atmosphere
resulting in strong dependence of the Mira
diameter observed in different absorption features (e.g.
Ireland et al. 2004; Woodruff et al. 2008, 2009; Zhao-Geisler et al.
2012).

  Models based on sophisticated treatment of radiative transfer have
become available in recent years (H{\"o}fner et al. 2003; Ireland et al.
2008 ({\tt CODEX1}), 2011 ({\tt CODEX2})). The {\tt CODEX} model
series, which are based upon a self-excited pulsation model for each
series, show that differences in position and strength of outward
travelling - or sometimes receding - shock fronts in different cycles
may lead to noticeable cycle-to-cycle differences of density and
temperature in upper layers and to differences
of spectral features formed in these
layers. Comparison of these models with observations of Miras were
published by Woodruff et al. (2009), Wittkowski et al. (2011),
Hillen et al. (2012).

  {\tt CODEX} models have so far been computed for 4 different sets of
parameters given by 4 non-pulsating ``parent stars''
(series R52, C50, C81, o54; see Tab.~1).
Solar element abundances (Z = 0.02; Grevesse et al. 1996) were
adopted for all series. Details of constructing pulsation models and
of computing atmospheric temperatures and spectra, based upon an
opacity-sampling treatment of absorption coefficients, are given in
Ireland et al. (2008, 2011). In this paper, we construct two
{\tt CODEX} model series with (i) ``mildly'' sub-solar metallicity and
(ii) ``mild'' S-type C/O abundance ratio.
In order to look for abundance-dependent differences,
we compare stratifications and colours of these models with
those of models of the o54 series which has, except for abundances,
almost identical parameters.

  We note that intermediate-period Mira variables
are most typically associated with the thick disk
(Groenewegen \& Blommaert 2006) which does not have solar abundances and,
in particular, has [Ti/Fe] enhancement of +0.2 and [O/Fe] and [C/Fe]
enhancements of +0.3 to +0.4 (e.g. Reddy et al. 2006).
Assuming a standard [$\alpha$/Fe]=0.0,
our chosen metallicity Z=0.02/3 corresponds to
[Ti/H]=-0.5, [O/H]=-0.5 and [C/H]=-0.5.
Given that Ti, O and C are the heavy elements that most significantly
influence the spectra of O-rich Mira variables, the metallicity
Z=0.02/3 can thus approximately represent a thin-disk star of
[Fe/H]=-0.5 as well as a thick-disk star with typical 
$\alpha$-enhancement of +0.3 and with [Fe/H]=-0.8.
This range is typical of stars kinematically associated with
the thick disk (e.g. Adibekyan et al. 2011, Cheng et al. 2012).
Of course, significant abundance variations are still expected throughout
the intermediate-period Mira variable population. This study attempts
to look for simple observational effects of these variations.

\section{Model parameters and descriptions}

  Inspection of the models presented in Ireland et al. (2011) shows that
the luminosity $L$, the $\tau_{\rm_{Ross}}$ = 1 Rosseland radius $R$,
and the positions and heights of atmospheric shock fronts can all
differ noticeably in different cycles. This leads to different effective
temperatures $T_{\rm eff} \propto (L/R^{2})^{1/4}$
and to different atmospheric temperature-density stratifications for
models at the same phase in different cycles. We shall investigate in
this study whether typical effects of moderately lower metallicity and
moderately higher S-type C/O ratio are strong enough to overcome
cycle-to-cycle differences which show up in standard spectral colours.
The o54 model series, assumed to have parameters close to $o$~Cet
(cf. Ireland et al. 2008 for details), is adopted as the
solar-abundance reference series in the present study.  The two
new model series are:

(i) The x54 series (see Tab.~1),
with Z = 0.02/3, is the low-metallicity counterpart
of the o54 series. The basic parent-star parameters,
the period
and the turbulent-velocity parameter of the x54 series are
almost identical to the o54 values.
And the mixing-length parameter (2.36) is somewhat lower than the
o54 value (3.5), which compensates for the lower metallicity in order to
match the observed period for given model luminosity and mass. The
lower mixing-length parameter represents less efficient convection,
increasing the stellar radius, while the lower metallicity lowers
opacity, making it easier for radiation to escape, and decreases the
stellar radius. A detailed discussion of these effects is given in
Ireland et al. (2008, 2011).

(ii) The s54 series (see Tab.~1)
represents the ``mild'' S-type counterpart of the
o54 series with a C enhancement of +0.274, resulting in [C/O]=0.9, and
with a Zr enhancement of +1.0 with respect to the solar values given in
Grevesse et al. (1996).
Since the grey pulsation model is hardly affected by such
``mild'' CNO changes ($T_{\rm eff}$ changes by only $\sim$10\,K for
fixed $M$, $L$ and $P$),
the pressure stratification of the original
o54 grey pulsation model was adopted for calculating the s54 non-grey
atmospheric temperature stratification and spectrum with the S-type
opacities.

  The behaviour of the x54 pulsation model series is shown in Fig.~1.
Two time intervals, covering 3 cycles, were selected for computing
atmospheric stratifications and they are marked as blue-shaded regions.
Basic parameters of the models of these 3 cycles, as well as
shock-front positions, are given in Tab.~2.

  The behaviour of the solar-abundance o54 pulsation model,
used for computing the S-type s54 series
in the present paper, is shown in Fig.~1 and in Tabs.~2 to 4
of Ireland et al. (2011). Three sub-intervals comprising 3 cycles
with different phase-dependent characteristics
($L$, $R$, $T_{\rm eff}$, shock front position/height)
were chosen for computing S-type atmospheric stratifications.
It turns out that Rosseland radii $R$ - and resulting effective
temperatures $T_{\rm eff}$ - at a given phase are almost identical
in s54 and o54 models except around phase 0.6 where s54 radii are
moderately smaller (by about 0.2 parent-star radii) and effective
temperatures are moderately higher (by about 200 K) than o54 values.

\begin{figure}
\centering
\includegraphics[width=0.8\columnwidth]{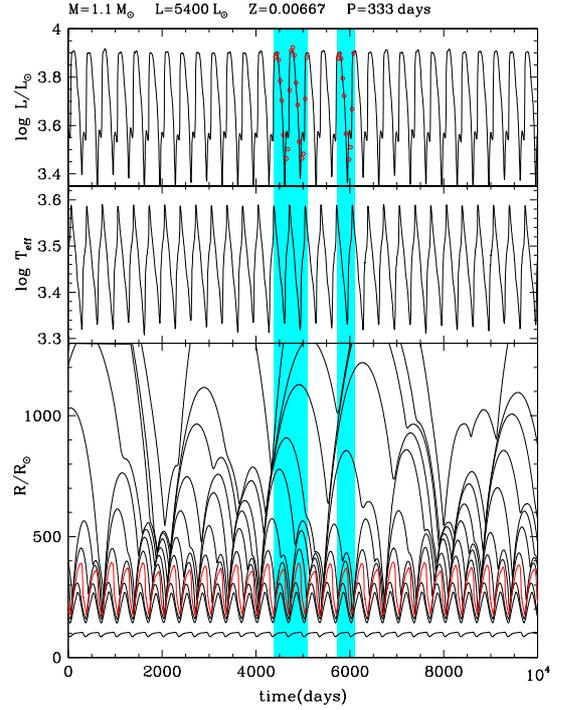}
\caption{
Luminosity $L$, effective temperature $T_{\rm eff}$
and radius $R$ of a representative selection of mass zones
plotted against time $t$ for the low-metallicity model series x54.
The red line in the $R$ panel shows the position of the
grey-approximation optical depth $\tau_g$=2/3. $T_{\rm eff}$
in the grey pulsation model is defined as the temperature
at $\tau_g$=2/3 and is close to the effective temperature
$\propto (L/R^2)^{1/4}$ of the non-grey atmospheric
stratification. The blue-shaded regions show the time intervals in
which models were selected for computing detailed atmospheres
(shown as red circles in the $L$ panel).
}
\label{fig1}
\end{figure}

\begin{figure}
\centering
\includegraphics[width=0.8\columnwidth]{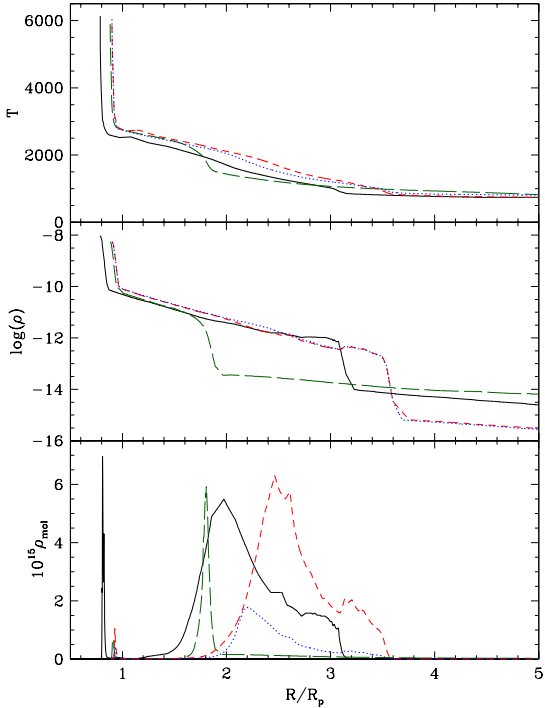}
\caption{
Temperature $T$ (upper), density $\rho$ (middle) and
density $\rho_{mol}$ of water molecules H$_2$O (lower panel)
as a function of radius $R$/R$_p$ for 4 phase 0.80 models:
metal-poor x54 series (black full line), solar-abundance o54 series
(green dashed, red short-dashed), S-type s54 series (blue dotted).
%
}
\label{fig2}
\end{figure}

\begin{figure*}
\centering
\includegraphics[width=0.6\textwidth]{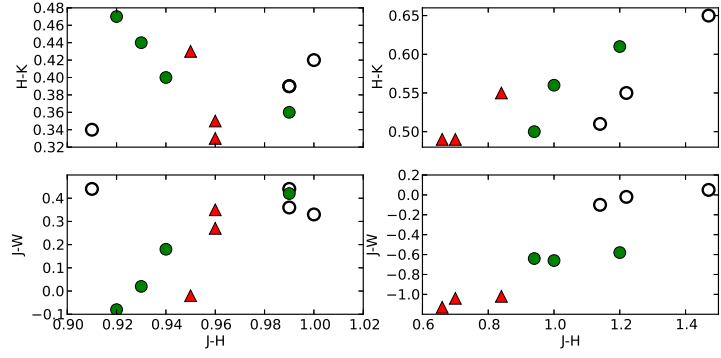}
\caption{
2-colour-diagrams of (H-K) (upper panel) and (J-W)
(lower panel) vs. (J-H). Left panel: phase 0.8 models of
the o54 series (filled green circles), the x54 series (filled red triangles)
and the s54 series (black circles).
Right panel: the same as the left panel but for phase 0.6 models.
}
\label{fig3}
\end{figure*}

\begin{table}
\caption{Parameters of 4 solar-abundance model series of
  Ireland et al. (2011) and the x54 and s54 model series
  presented here: mass $M$, luminosity $L$,
  metallicity $Z$, mixing-length parameter $\alpha_m$, turbulent
  viscosity parameter $\alpha_\nu$,
  parent-star radius $R_p$, period $P$.
  The s54 model series has solar $Z$=0.02 with increased
  S-type C and Zr abundances (see text). The full model output of all
  models is provided at http://www.mso.anu.edu.au/$\sim$mireland/codex/}
\begin{tabular}{llllllll}
\hline
Name & M & L & Z & $\alpha_m$ &$\alpha_\nu$& $R_p$ & $P$\\
\hline
{\tt R52} & 1.1  & 5200 & 0.02   & 3.5  & 0.25 & 209 & 307 \\
{\tt C50} & 1.35 & 5050 & 0.02   & 2.0  & 0.24 & 291 & 427 \\
{\tt C81} & 1.35 & 8160 & 0.02   & 3.5  & 0.32 & 278 & 430 \\
{\tt o54} & 1.1  & 5400 & 0.02   & 3.5  & 0.25 & 216 & 330 \\
{\tt x54} & 1.1  & 5400 & 0.0067 & 2.36 & 0.25 & 224 & 333 \\
{\tt s54} & 1.1  & 5400 & -      & 3.5  & 0.25 & 216 & 330 \\
\hline
\end{tabular}
\end{table}

\begin{table}
\caption{Parameters of the x54 models including positions of
shock fronts. (The non-pulsating "parent star" of this model
series has mass $M$ = 1.1 M$_\odot$, luminosity $L_p$ =
5400 L$_\odot$ and radius $R_p$ = 224 R$_\odot$).}
\resizebox{1.0\columnwidth}{!} {
\begin{tabular}{lcccccccc}
\hline
Model &  Phase & L & R & T$_{\rm eff}$ & S1 & S2 & S3 & S4  \\
      &        & ($L_\odot$) & ($R_p$) & (K) & ($R_p$) & ($R_p$)
                                             & ($R_p$) & ($R_p$) \\
\hline
237070&  1.00&  7652&    1.00&  3601&    2.72&   1.05&       &        \\
237210&  1.10&  7909&    1.19&  3332&    2.69&   1.32&       &        \\
237320&  1.20&  7442&    1.29&  3153&    2.58&   1.54&       &        \\
237420&  1.30&  6097&    1.32&  2971&    2.50&   1.68&       &        \\
237510&  1.40&  5049&    1.28&  2875&    2.40&   1.76&       &        \\
237570&  1.49&  3637&    1.21&  2722&    2.21&   1.81&       &        \\
237630&  1.60&  1648&    1.28&  2172&    2.00&   1.79&   0.99&        \\
237810&  1.70&  2906&    0.87&  3033&    $=>$&   1.70&   0.88&        \\
238030&  1.80&  3177&    0.80&  3245&        &   1.61&   0.81&        \\
238360&  1.90&  5582&    0.78&  3780&        &   1.40&   0.79&        \\
238940&  2.00&  8110&    1.02&  3634&        &   $=>$&   1.13&        \\
239080&  2.10&  8366&    1.20&  3365&        &       &   1.61&        \\
239150&  2.20&  7766&    1.30&  3174&        &       &   1.99&        \\
239190&  2.31&  5979&    1.33&  2946&        &       &   2.31&        \\
239210&  2.41&  4851&    1.29&  2840&        &       &   2.55&        \\
239230&  2.51&  3413&    1.21&  2679&        &       &   2.74&        \\
239270&  2.60&  2007&    1.26&  2298&        &       &   2.87&   1.01 \\
239420&  2.70&  2909&    0.88&  3020&        &       &   2.99&   0.88 \\
239550&  2.80&  3044&    0.80&  3205&        &       &   3.09&   0.81 \\
239690&  2.90&  5124&    0.76&  3735&        &       &   3.17&   0.77 \\
239880&  3.00&  7628&    1.00&  3605&        &       &   3.23&   1.04 \\
      &      &      &        &      &        &       &       &        \\
242850&  5.00&  7554&    0.99&  3621&    3.61&   1.02&       &        \\
243030&  5.10&  7892&    1.17&  3357&    3.74&   1.29&       &        \\
243170&  5.20&  7516&    1.28&  3172&    3.82&   1.52&       &        \\
243270&  5.30&  6234&    1.32&  2989&    3.91&   1.71&       &        \\
243330&  5.39&  5238&    1.29&  2894&    3.98&   1.83&       &        \\
243380&  5.50&  3680&    1.21&  2733&    4.04&   1.91&       &        \\
243420&  5.60&  1850&    1.25&  2264&    4.05&   1.93&       &        \\
243590&  5.70&  2877&        &      &    4.08&   1.89&   0.88&        \\
243780&  5.80&  3238&    0.80&  3248&    4.10&   1.80&   0.81&        \\
244000&  5.90&  4654&    0.77&  3634&    4.05&   1.66&   0.78&        \\
244500&  6.00&  7925&    1.00&  3641&    4.05&   1.47&   1.06&        \\
\hline
\end{tabular}
}
\end{table}

\section{Abundance effects on temperature stratifications}

  Comparison of the effective temperatures of the metal-poor x54
models (Tab.~2) with those of o54 models
(Tabs.~2 to 4 of Ireland et al. 2011)
shows that, despite occasionally significant luminosity
differences at some phases, effective temperatures at a given
phase are very similar in both series. This means that
$T_{\rm eff} \propto (L/R^{2})^{1/4}$
as a function of phase hardly changes with moderately decreasing
metallicity. Only around phase 0.4 can a very modest trend towards
somewhat higher x54 effective temperatures
(about 100 to 200K) be seen.

  When comparing the S-type s54 models with the o54 models, we are
dealing with strictly differential effects caused by the CNO abundance
differences between s54 and o54 (since the s54 models are based upon
the pressure stratifications of the corresponding models
in the o54 series). At most phases, no systematic difference
between s54 and o54 effective temperatures can be found, except around
phase 0.6 where the s54 model in a given cycle tends to be hotter
by about 200 K than its o54 counterpart, but cycle-to-cycle
differences still lead to marginal $T_{\rm eff}$ overlaps.

  Since the temperature stratification of the
middle to upper atmosphere of
a Mira variable tends to depend, at a given phase, more on the
positions of shock fronts (and their associated density
stratification) than on the exact value of $T_{\rm eff}$, we
have to accept that modest cycle-dependent changes of
$T_{\rm eff} \propto (L/R^{2})^{1/4}$ cannot be safely derived
from observations of spectral
features formed in these layers.

  Figure~2 shows stratifications of temperature, density and H$_2$O
molecular density at phase 0.8 for a typical model of the x54
series, for 2 typical models (of different cycles) of the o54
series, and for the s54 counterpart of one of these o54 models.
Cycle-to-cycle differences of the atmospheric stratification
as shown, e.g., in Fig.~2 for two o54 models lead to significant
spectrum differences (cf. Ireland et al. 2011). These interfere with
differences caused by changing element abundances and may, of course,
affect the assignment of colours to abundances.

\section{Abundance effects on colours}

  From the findings discussed in Sect.~3, we would not expect to
see really significant abundance-related spectrum differences between
our model series. This is confirmed by systematic inspection of standard
infrared colour indices $J-H$, $J-K$ and $H-K$ in all models which were
investigated in this study. We added a set of 3 colour indices that
combines $JHK$ with a ``water colour'' $W$, defined by the flux in
a narrow water-dominated bandpass (rectangular profile) centered at
1.45 $\mu$m with full width
0.087 $\mu$m. These colour indices, too, do not show clear
abundance effects. (Inspection of fluxes in the $I$ bandpass and
in the optical $R$ and $V$ bandpasses shows still more pronounced
cycle-to-cycle differences so these colours could be a priori
excluded as useful abundance indicators.)

  The left panel of Fig.~3 shows 2-colour-diagrams ($H-K$) vs. ($J-H$) and
($J-W$) vs. ($J-H$) for all phase 0.8 models of this study.  The
stratifications of four of these models are shown in Fig.~2. No clear
separation of x54, o54 and s54 models can be seen in this figure:
cycle-to-cycle stratification differences clearly dominate
abundance effects. This behaviour holds for the models at all
pulsation phases except for models near phase 0.6 where some
systematic differences between models of different abundances can
be seen in the 2-colour diagrams (right panel of Fig.~3).
Inspection of Tab.~1 and of Tabs.~2 to 4 of Ireland et al.(2011)
shows that the effective temperature at near-minimum phase 0.6 is
substantially lower than at neighbouring phases so that the influence
of relevant molecules (especially water) upon the stratification
and the spectrum is noticeably higher than at slightly later and,
in particular, slightly earlier phases (where phase details
somewhat depend on specific stellar parameters: cf. Tabs.~2 to 8
of Ireland et al. (2011)).
A careful observational campaign covering phases near minimum
in small phase steps of, e.g., 0.05 would be needed to clearly see
this effect in real stars. It is also important that, when making
conclusions about abundances from a model series, this series must
have parameters that match broadband colours, pulsation period and
amplitude.

  We also inspected the phase and cycle dependence of other
water features, e.g. in the K band range, and found similar, but
sometimes less pronounced and less obvious, abundance effects as
in the W band. Colours using this band appear to be preferable
abundance indicators.
  
  One also has to be aware that the {\tt CODEX} models
have so far been compared to only a few stars,
with more or less accurately known stellar parameters, at
selected phases/cycles in selected wavelength ranges. Woodruff
et al. (2009) compare fluxes and diameters of 3 stars
(W~Hya, R~Leo, $o$~Cet; 2 phases) in the 1.0 to 4.0 $\mu$m range
with o54 models and find satisfactory agreement with some
differences at longer wavelengths. Wittkowski et al. (2011)
compare fluxes and visibilities/diameters of 4 stars
(R~Cnc, X~Hya, W~Vel, RW~Vel; 1 phase) in the 1.9 to 2.5 $\mu$m
range with models (C50, o54, C81, o54, respectively) and find
satisfactory agreement with only moderate differences, although
the measured closure phases indicate noticeable deviations from
spherical symmetry not considered in the spherical {\tt CODEX}
models. Hillen et al. (2012) compare fluxes and visibilities
of TU~And in the 2.0 to 2.4 $\mu$m range for a large
number of phases of 8 successive cycles with predictions
of R52 (and o54) models. The TU~And observations are well reproduced
by the models at near-minimum phases but suggest that this star's
atmosphere is more extended than the model atmosphere at
near-maximum phases. If this indicates a general model problem,
rather than imperfect stellar vs. model parameter assignment,
the present study might underestimate abundance effects on colours
near maximum phases.

  It might appear astonishing at first glance that ``mild'' changes
of abundances, in particular lowering metallicity, results in only
fairly small effects upon colours which can hardly be disentangled
from cycle-to-cycle differences of the atmospheric structure. One
has to be aware, however, that hydrogen is not a major absorber
in the upper layers of such very cool atmospheres. Decreasing
the metal-to-hydrogen ratio affects essentially 2 absorbers,
H$^{-}$ and H$_2$O, where electrons in H$^{-}$ come from metals,
and the resulting changes of the temperature stratification
(cf. Fig.~2) and of the emitted spectrum are very modest.

\section{Concluding remarks}

  We computed two model series which allow the study of the effects
of ``mild'' deviations from solar element abundances upon the
atmospheric stratification and standard spectral colours of
typical ($o$~Cet-like) Mira variables.  The x54 model series is
the $\frac{1}{3}$ metallicity counterpart of the solar-metallicity
o54 series discussed by Ireland et al. (2008, 2011),
while the s54 model series is the S-type counterpart
of the o54 series. Both model series show noticeable
stratification and spectral-colour differences from
the o54 models.

  It turns out, however, that such abundance effects are
readily masked by significant cycle-to-cycle differences
at most phases. Also, in real stars, one needs to 
consider in addition the effects caused by modest differences
of stellar parameters, so one has to conclude that
``mild'' deviations from solar element abundances will barely
be detectable in the structure of the stellar atmosphere and the
broadband emitted spectral flux. This confirms the findings of
Scholz \& Wood (2004) who infer, based on a series of less
elaborated models, that there is no easy way to determine
the metallicity of an M-type Mira field star for moderate
deviations (factor of the order of 2) from solar metallicity.
We suggest that future observational campaigns should both
focus on phase-dependent measurements in the W water filter
and on high-spectral-resolution bandpasses that include a continuum
(e.g. J and H bands) and where the classical method of
line analysis can be attempted.


\label{lastpage}


\begin{thebibliography}{}
%
\bibitem{}
Adibekyan, V.~Zh., Santos, N.~C., Sousa, S.~G., \& Israelian, G.
2011, A\&A, 535, L11
%
\bibitem{}
Cheng, J.~Y., Rockosi, C.~M., Morrison, H.~L., Sch{\"o}nrich, R.~A.,
  Lee, Y.~S., Beers, T.~C., Bizyaev, D., Pan, K.,\& Schneider, D.~P.
  2012, ApJ, 746, 149
%
\bibitem{}
Fox, M.~W., Wood, P.~R.,\& Dopita, M.~A. 1984, ApJ, 286, 337
%
\bibitem{}
Grevesse, N., Noels, A.,\& Sauval, A.~J. 1996, in: Holt, S.~S.,
  Sonneborn, G. eds., ASP Conf. Ser. Vol. 99, Cosmic Abundances, p.117
%
\bibitem{}
Groenewegen, M.A.T.,\& Blommaert, J.A.D.L. 2006, Mem. Soc. Astron. Ital.,
  77, 81
%
\bibitem{}
Hillen, M., Verhoelst, T., Degroote, P., Acke, B.,\& Van Winckel, H. 2012,
  A\&A, 538, L6
%
\bibitem{}
H{\"o}fner, S., Gautschy-Loidl, R., Aringer, B.,\& J{\o}rgensen, U.~G. 2003,
  A\&A, 399, 589
%
\bibitem{}
Ireland, M.~J., Scholz, M.,\& Wood, P.~R. 2008, MNRAS, 391, 1994
%
\bibitem{}
Ireland, M.~J., Scholz, M.,\& Wood, P.~R. 2011, MNRAS, 418, 114
%
\bibitem{}
Ireland, M.~J., Tuthill, P.~G., Bedding, T.~R., Robertson, J.~G.,\&
  Jacob, A.~P. 2004, MNRAS, 350, 365
%
\bibitem{}
Reddy, B.~E., Lambert, D.~L.,\& Allende Prieto, C. 2006, MNRAS, 367, 1329
%
\bibitem{}
Richter, He.,\& Wood, P.~R. 2001, A\&A, 369, 1027
%
\bibitem{}
Scholz, M.,\& Wood, P.~R. 2004, in: Kurtz D.~W., Pollard K. eds., IAU Coll.
  193, ASP Conf. Ser. Vol. 310, Variable Stars in the Local Group, p.313
%
\bibitem{}
Wittkowski, M., Boboltz, D.~A., Ireland, M.~J., Karovicova, I., Ohnaka, K.,
  Scholz, M., Van Wyk, F., Whitelock, P., Wood, P.~R.,\& Zijlstra, A.~A 2011,
  A\&A, 532, L7
%
\bibitem{}
Woodruff, H.~C., Tuthill, P.~G., Monnier, J.~D., Ireland, M.~J.,
  Bedding, T.~R., Lacour, S., Danchi, W.~C.,\& Scholz, M. 2008, ApJ, 673, 418
%
\bibitem{}
Woodruff, H.~C., Ireland, M.~J., Tuthill, P.~G., Monnier, J.~D.,
  Bedding, T.~R., Danchi, W.~C., Scholz, M., Townes, C.~H.,\& Wood, P.~R. 2009,
  ApJ, 691, 1328
%
\bibitem{}
Zhao-Geisler, R., Quirrenbach, A., K{\"o}hler, R.,\& Lopez, B. 2012,
  A\&A, 545, A56
%
\end{thebibliography}
\end{document}